\documentclass[
prc,%
10pt,%
final,%
notitlepage,%
oneside,%
twocolumn,%
nobibnotes,%
nofootinbib,
superscriptaddress,%
floatfix,%
floatfix,%
showkeys,%
showpacs]%
{revtex4}
\usepackage{color}
\usepackage{amsfonts}
\usepackage{amsbsy}
\usepackage{mathrsfs}
\usepackage{graphicx}
\def\lsim{\mathrel{\rlap{
\lower4pt\hbox{\hskip-3pt$\sim$}}
    \raise1pt\hbox{$<$}}}     
\def\gsim{\mathrel{\rlap{
\lower4pt\hbox{\hskip-3pt$\sim$}}
    \raise1pt\hbox{$>$}}}     
\def\scr#1{\mbox{\scriptsize #1}}
\begin{document}
\title{	
Estimates of hyperon polarization in heavy-ion collisions at collision energies $\sqrt{s_{NN}}=$4--40 GeV
} 
\author{Yu. B. Ivanov}\thanks{e-mail: yivanov@theor.jinr.ru}
\affiliation{Bogoliubov Laboratory for Theoretical Physics, 
Joint Institute for Nuclear Research, Dubna 141980, Russia}
\affiliation{National Research Nuclear University "MEPhI", 
Moscow 115409, Russia}
\affiliation{National Research Centre "Kurchatov Institute",  Moscow 123182, Russia} 
\author{V. D. Toneev}
\affiliation{Bogoliubov Laboratory for Theoretical Physics, 
Joint Institute for Nuclear Research, Dubna 141980, Russia}
\author{A. A. Soldatov}
\affiliation{National Research Nuclear University "MEPhI",
Moscow 115409, Russia}
\begin{abstract}
Global polarization of $\Lambda$ 
and $\bar{\Lambda}$ hyperons in Au+Au collisions at collision energies $\sqrt{s_{NN}}=$ 4-40 GeV
in the midrapidity region and total polarization, 
i.e. averaged over all rapidities, are studied within the scope of   
the thermodynamical approach. 
The relevant vorticity is simulated within the model of the three-fluid dynamics (3FD). 
It is found that 
the performed rough estimate of the global midrapidity polarization
quite satisfactorily reproduces the experimental STAR data on the  
polarization, especially its collision-energy dependence. 
The total polarization increases with the collision energy rise,  
which is in contrast to the decrease of the midrapidity polarization. 
This suggests that at high collision energies
the polarization reaches high values in fragmentation regions.
\end{abstract}
\maketitle

\section{Introduction}

Huge global angular momentum is generated in non-central heavy-ion collisions at high energies 
that can be partially transformed into spin alignment of constituents \cite{Liang:2004ph,Betz:2007kg,Gao:2007bc}. The latter can be measured by the polarization of hyperons and vector mesons.
Global polarization of $\Lambda$ and $\bar{\Lambda}$ hyperons  was measured \cite{STAR:2017ckg} by the STAR experiment in the energy range of 
the Beam Energy Scan (BES) program at the  Relativistic Heavy Ion Collider (RHIC)
at Brookhaven. It was measured in the midrapidity region of colliding nuclei. 
The measured polarization is 
generally reproduced within the hydrodynamic \cite{Karpenko:2016jyx,Xie:2017upb} and 
kinetic \cite{Li:2017slc,Sun:2017xhx,Kolomeitsev:2018svb,Wei:2018zfb,Shi:2017wpk} model calculations
based on the thermodynamics  in the hadronic phase 
\cite{Becattini:2013fla,Fang:2016vpj,Becattini:2016gvu}, as well as within an alternative 
approach directly based on the axial vortical effect 
\cite{Rogachevsky:2010ys,Gao:2012ix,Sorin:2016smp} within the quark-gluon string transport model 
 \cite{Baznat:2017jfj}. 
The axial vortical effect is associated with axial-vector current 
induced by vorticity. This current implies that the right (left)-handed
fermions move parallel (opposite) to the direction of vorticity. As the momentum of a right (left)-handed massless fermion is parallel (opposite) to its spin, all spins become
parallel to the direction of vorticity, i.e. aligned.

In the present paper we estimate the global polarization of $\Lambda$ 
and $\bar{\Lambda}$ 
hyperons in Au+Au collisions based on the thermodynamical approach 
\cite{Becattini:2013fla,Fang:2016vpj,Becattini:2016gvu}.    
The relevant vorticity is simulated within the model of the three-fluid dynamics (3FD) \cite{3FD}.  
We perform a collision-energy scan in the energy range of the Facility for Antiproton and Ion Research (FAIR) in Darmstadt \cite{Friman:2011zz}, the Nuclotron based
Ion Collider fAcility (NICA) in Dubna \cite{Kekelidze:2017ghu} and BES at RHIC.

The 3FD model describes of the major part of bulk
observables: the baryon stopping \cite{Ivanov:2013wha,Ivanov:2012bh}, 
yields of different hadrons, their rapidity and transverse momentum
distributions \cite{Ivanov:2013yqa,Ivanov:2013yla}, and also  
the elliptic \cite{Ivanov:2014zqa} 
and directed \cite{Konchakovski:2014gda} flow. 
It also reproduces \cite{Ivanov:2018vpw} recent STAR data on bulk observables \cite{Adamczyk:2017iwn}. 

The question we address in this paper is whether the 3FD model is able to reproduce 
the observed global midrapidity polarization without any additional adjustment of the model parameters.  
In other words, are the bulk and flow properties of the produced matter  
internally interconnected with its polarization? Based on this analysis we make predictions for  
the global midrapidity polarization in the FAIR-NICA energy range.

We also address the question why does the observed global polarization of 
hyperons in the midrapidity region 
drop with the collision energy rise while the total angular momentum 
accumulated in the system substantially increases at the same time? To this end, we 
also estimate total polarization of hyperons, i.e. the mean global polarization over 
all rapidities.

\section{The 3FD Model}
\label{Model}

The  3FD model takes into account 
a finite stopping power resulting in counterstreaming 
of leading baryon-rich matter at the early stage of nuclear collisions 
\cite{3FD}. This 
nonequilibrium stage 
is modeled by means of two counterstreaming baryon-rich fluids 
initially associated with constituent nucleons of the projectile
(p) and target (t) nuclei. 
Later on these  fluids may consist
of any type of hadrons and/or partons (quarks and gluons),
rather than only nucleons.
Newly produced particles, dominantly
populating the midrapidity region, are associated with a fireball
(f) fluid.
These fluids are governed by conventional hydrodynamic equations 
coupled by friction terms in the right-hand sides of the Euler equations. 
The friction results in energy--momentum loss of the 
baryon-rich fluids. A part of this
loss is transformed into thermal excitation of these fluids, while another part 
leads to formation of the fireball fluid.
Thus, the 3FD approximation is a minimal way to implement the early-stage nonequilibrium 
of the produced strongly-interacting matter at high collision energies.

The physical input of the present 3FD calculations is described in
Ref.~\cite{Ivanov:2013wha}. 
Three different 
equations of state (EoS's) were used in simulations of Refs. 
\cite{Ivanov:2013wha,Ivanov:2012bh,Ivanov:2013yqa,Ivanov:2013yla,Ivanov:2014zqa,Konchakovski:2014gda,Ivanov:2018vpw}: a purely hadronic EoS \cite{gasEOS}  
and two versions of the EoS with the   deconfinement
 transition \cite{Toneev06}, i.e. a first-order phase transition  
and a crossover one. 
The friction between the fluids in the hadronic phase was estimated in Ref. \cite{Sat90}
based on experimental proton-proton cross sections. This friction is implemented 
in the 3FD simulations of the hadronic phase. 
There are no estimates of this friction in the quark-gluon 
phase (QGP). Therefore, the friction in the QGP was fitted for each EoS to reproduce
the observed stopping power, 
see  Ref. \cite{Ivanov:2013wha} for details.
In the present paper  
only the first-order-phase-transition (1st-order-tr.) and crossover EoS's 
are used as 
the most relevant to various observables.

\begin{figure}[htb]
\includegraphics[width=8.2cm]{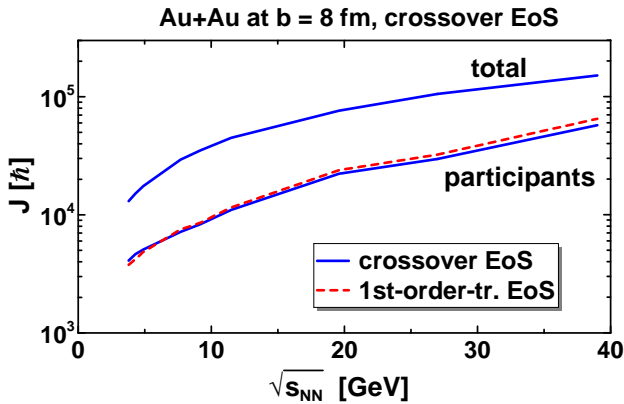}
 \caption{(Color online)
The total angular momentum (conserved quantity) and  
the angular momentum accumulated in the participant 
region 
in semi-central ($b=$ 8 fm) Au+Au collision as functions of $\sqrt{s_{NN}}$. 
Calculations are done with the 1st-order-transition and crossover EoS's. 
}
\label{fig1}
\end{figure}

Total angular momentum is conserved with an accuracy of 1\%
in the 3FD simulations. The angular momentum is defined as 
   \begin{eqnarray}
   \label{J-tot}
J = \int d^3 x \sum_{\alpha=\rm{p,t,f}} (z\; T^{\alpha}_{10} - x\; T^{\alpha}_{30}).    
   \end{eqnarray}
where $T^{\alpha}_{\mu\nu}$ is the energy-momentum tensor of the $\alpha$(=p,t,f) fluid  
and has the conventional hydrodynamical form, $z$ is the beam axis, 
$(x,z)$ is the reaction plane of the colliding nuclei. 
The total angular momentum, $J_{\rm{total}}$, in semi-central (impact parameter $b=$ 8 fm) 
Au+Au collision as function of 
collision energy $\sqrt{s_{NN}}$ is presented in Fig. \ref{fig1}. 
It is independent of the used EoS. 
For the $J_{\rm{total}}$ calculation the integration in Eq. (\ref{J-tot}) runs over 
the whole system. As seen, $J_{\rm{total}}$ rapidly rises with the collision energy. 
Only a part of the total angular momentum is accumulated in the participant 
region. Figure \ref{fig1} also displays the angular momentum accumulated in the participant 
region, i.e. in the overlap region of the interacting fluids. 
As seen from Fig. \ref{fig1}, 25--30\% of the total angular momentum is deposited into 
participant matter in the Au+Au collisions at $b=$ 8 fm.

The participant angular momentum rises with time because 
the overlap region of the interacting fluids
increases in the course of the expansion stage and includes 
more and more former spectators.  
Therefore, the participant angular momentum depends, though weakly, on the EoS. 
Figure \ref{fig1} presents the participant angular momenta at the 
``freeze-out'' instant of time in the c.m. frame of colliding nuclei, 
i.e. when average energy density throughout the participant region falls
to the freeze-out value of $\varepsilon_{\rm frz}$ = 0.4 GeV/fm$^3$. 
This is a kind of an illustrative freeze-out. 
In actual calculations of observables a differential, i.e. cell-by-cell,  
freeze-out is implemented \cite{Russkikh:2006aa}. 
The freeze-out occurs when the local energy density drops down to the 
freeze-out value $\varepsilon_{\rm frz}$.

\section{Vorticity in the 3FD model}
\label{Results}

A so-called thermal vorticity is defined as 
   \begin{eqnarray}
   \label{therm.vort.}
   \varpi_{\mu\nu} = \frac{1}{2}
   (\partial_{\nu} \hat{\beta}_{\mu} - \partial_{\mu} \hat{\beta}_{\nu}), 
   \end{eqnarray}
which is dimensionless.
Here 
$\hat{\beta}_{\mu}=\hbar\beta_{\mu}$,  $\beta_{\mu}=u_{\nu}/T$, 
$u_{\mu}$ is collective local four-velocity of the matter,  and
$T$ is local temperature.  
In the thermodynamical approach \cite{Becattini:2013fla,Fang:2016vpj,Becattini:2016gvu}
in the leading order in the thermal vorticity 
it is directly related to 
the mean spin vector of spin 1/2 particles with four-momentum $p$, 
produced around point $x$ on freeze-out hypersurface 
   \begin{eqnarray}
\label{xp-pol}
 S^\mu(x,p)
 =\frac{1}{8m}     [1-n_F(x,p)] \: p_\sigma \epsilon^{\mu\nu\rho\sigma} 
  \varpi_{\rho\nu}(x) 
   \end{eqnarray}
where $n_F(x,p)$ is the Fermi-Dirac distribution function and $m$ is mass of the 
considered particle. 
To calculate the
relativistic mean spin vector of a given particle species
with given momentum, the above expression should be integrated
over the freeze-out hypersurface.

Unlike the conventional hydrodynamics,  
the system is characterized by three hydrodynamical
velocities, $u^\mu_a$ ($a$ = p, t and f), in the 3FD model. 
The counterstreaming of
the p and t fluids takes place only at the initial stage of
the nuclear collision  
that lasts from $\sim$ 5 fm/c at $\sqrt{s_{NN}}=$ 5 GeV \cite{Ivanov:2017dff}
to $\sim$ 1 fm/c at collision energy of 39 GeV \cite{Ivanov:2017xee}. 
At later stages the baryon-rich (p and t) fluids have already 
either partially passed though each other or 
partially stopped and unified in the central region. 
At lower collision energies, like those of NICA and FAIR, the contribution 
the f-fluid into various quantities, in particular into the vorticity \cite{Ivanov:2017dff}, 
is small compared with that of the baryon-rich (p and t) fluids. 
At higher BES RHIC energies the f-fluid contributions become 
comparable with those of the baryon-rich (p and t) fluids. 
The f-fluid also is entrained by the the unified baryon-rich fluid 
but is not that well unified with the latter, thus keeping its 
identity even after the initial thermalization/unification of the baryon-rich fluids. 
The local baryon-fireball relative velocity is
small but not negligible even at the freeze-out stage \cite{Ivanov:2018eej}. 
In particular, the friction between
the baryon-rich and net-baryon-free fluids is the
only source of dissipation at the expansion stage. 
Therefore, after the initial thermalization stage the system is characterized 
 by two hydrodynamical velocities, $u^\mu_{\rm B}$ and $u^\mu_{\rm f}$, 
and two temperatures, $T_{\rm B}$ and $T_{\rm f}$, corresponding to the 
unified baryon-rich (B) and fireball (f) fluids.

As a result the system is characterized by 
two sets of the  vorticity related to these baryon-rich   
and baryon-free fluids, 
$\varpi_{\mu\nu}^{\rm B}$ and $\varpi_{\mu\nu}^{\rm f}$, respectively, 
which are defined in terms of their velocities and temperatures. 
We consider a proper-energy-density weighted  vorticity which allows us 
to suppress contributions of regions of low-density matter.  
It is appropriate because production of (anti)hyperons under consideration 
dominantly takes place in highly excited regions of the system. 
We also make sum of vorticities of the baryon-rich and baryon-free fluids 
with the weights of their energy densities, and thus define a single 
quantity responsible for the particle polarization 
%
   \begin{eqnarray}
   \label{en.av.rel.B-vort}
   \widetilde{\varpi}_{\mu\nu} ({\bf x},t) 
   =  
   \frac{\varpi_{\mu\nu}^{\rm B}({\bf x},t) \varepsilon_{\rm B} ({\bf x},t)
   + 
   \varpi_{\mu\nu}^{\rm f}({\bf x},t)  \varepsilon_{\rm f} ({\bf x},t)}{ 
 \varepsilon ({\bf x},t)}   
   \end{eqnarray}
where $\varepsilon_{\rm B}$ and $\varepsilon_{\rm f}$ are the proper energy densities of the 
the baryon-rich and baryon-free fluids, respectively. 
The proper energy density of all three fluids in their combined local rest frame, $\varepsilon$,  
is 
\begin{eqnarray}
\label{eps_tot}
\varepsilon = u_\mu T^{\mu\nu} u_\nu. 
\end{eqnarray}
where 
$T^{\mu\nu} \equiv
T^{\mu\nu}_{\scr p} + T^{\mu\nu}_{\scr t} + T^{\mu\nu}_{\scr f}$
is the total energy--momentum tensor
being the sum of conventional hydrodynamical energy--momentum tensors of separate fluids, and
the total collective 4-velocity of the matter is 
\begin{eqnarray}
\label{u-tot}
u^\mu = u_\nu T^{\mu\nu}/(u_\lambda T^{\lambda\nu} u_\nu). 
\end{eqnarray}
However, because of almost perfect unification of the baryon-rich fluids 
and small local baryon-fireball relative velocities \cite{Ivanov:2017xee}, 
at the later stages of the collision
a very good approximation for $\varepsilon$ is just 
\begin{eqnarray}
\label{eps-tot-appr}
\varepsilon \simeq  \varepsilon_{\scr B} + \varepsilon_{\scr f}. 
\end{eqnarray}

A quantitative comparison of the thermal vorticity 
in semi-central ($b=$ 8 fm) Au+Au collisions at different  
collision energies $\sqrt{s_{NN}}$ 
is performed in terms of average thermal 
vorticity  of the composed matter 
[Eq.   (\ref{en.av.rel.B-vort})] also
averaged over coordinate ($x$) space 
with the weight of the proper energy density  
   \begin{eqnarray}
   \label{en.av.therm.B-vort-T}
  \langle \varpi_{\mu\nu} (t) \rangle
  &=& \int
  d^3 x \;
   [\varpi_{\mu\nu}^{\rm B}({\bf x},t)\;\varepsilon_{\rm B} ({\bf x},t) 
   \cr
   &+&\vphantom{\int dV}\varpi_{\mu\nu}^{\rm f}({\bf x},t)\;\varepsilon_{\rm f}({\bf x},t)]
 \Big/ \langle \varepsilon (t)\rangle
   \end{eqnarray}
where average energy density is
   \begin{eqnarray}
   \label{B-en.av.}
\langle \varepsilon (t) \rangle = 
\int d^3 x  \; \varepsilon ({\bf x},t) 
\Big/
\int \theta[\varepsilon ({\bf x},t)] d^3 x 
   \end{eqnarray}
with $\theta(x)$ being equal to 1 for $x>0$ and 0 otherwise.  
This averaging is performed over two different space regions:
\\
{\bf (a)} Over central slab, 
$|x| < R-b/2$, $|y| < R-b/2$
and $|z| < R/\gamma_{cm}$, 
where $R$ is the radius of the Au nucleus, $b$ is the impact parameter and 
$\gamma_{cm}$ is the Lorentz
factor associated with the initial nuclear motion along the beam ($z$) axis 
in the c.m. frame.
This central central layer includes 
the whole participant region in the transverse direction. 
The data from this central slab are used to imitate 
the midrapidity global polarization. 
\\
{\bf (b)}
Over the whole participant region system, which is 
restricted by the condition $T >$ 100 MeV. This condition first of all is related to 
the baryon-rich fluid because the temperature of the 
produced f-fluid is always high. 
The temperature gradients and hence 
the thermal vorticity reach very high values at the spectator-participant border, 
where the temperature itself is not that high.  
At the same time, the $\Lambda$ hyperons are efficiently produced only
from the hottest regions of the system. 
Therefore,  
keeping in mind application to the $\Lambda$ polarization, 
we apply this temperature constraint. 
The temperature is always high in the above discussed central slab, 
that makes this constraint unnecessary.

In order to keep all the matter in the 
consideration, conventional local 3FD freeze-out was turned off because it 
removes the frozen out matter from the hydrodynamical evolution \cite{Russkikh:2006aa}.
Nevertheless, we do apply 
a simplified freeze-out, that has been already mentioned in the end of the previous section. 
This is an isochronous freeze-out similar to that used in Refs. \cite{Karpenko:2016jyx,Xie:2017upb}. 
The system is frozen out at the time instant $t_{\rm frz}$ when 
\\
{\bf (a)} the average energy density in the central slab, 
$\langle \varepsilon (t) \rangle_{\rm slab} $, decreases to its  
freeze-out value $\varepsilon_{\rm frz}$ = 0.4 GeV/fm$^3$, or  
\\
{\bf (b)}
the average energy density in the whole participant region, 
$\langle \varepsilon (t) \rangle_{\rm total} $, decreases to its  
freeze-out value $\varepsilon_{\rm frz}$. 
\\
The freeze-out in the central slab of the system of colliding nuclei is used to imitate 
the midrapidity global polarization, 
while that in the whole participant region is used to estimate the total\footnote{
to distinguish it from the global one at the midrapidity} 
polarization that also includes averaging over all rapidities.

%
\begin{figure}[bht]
\includegraphics[width=7.cm]{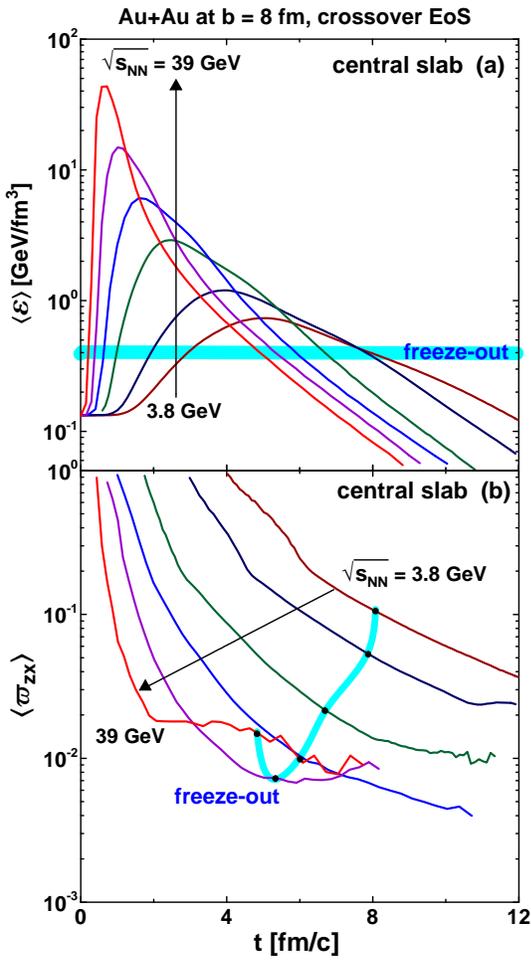}
 \caption{(Color online)
Time evolution of \\
(a) 
the proper energy density averaged over the central slab 
in the semi-central ($b=$ 8 fm) Au+Au collision at various $\sqrt{s_{NN}}=$
3.8, 4.9, 7.7, 11.5, 19.6, 39 GeV, 
the cyan band is placed at the freeze-out energy density $\varepsilon=$ 0.4 GeV/fm$^3$; 
\\
(b) 
the proper-energy-density-weighted thermal $zx$ vorticity averaged over the central slab, 
the cyan band indicates the freeze-out, corresponding to $\varepsilon=$ 0.4 GeV/fm$^3$ 
band in panel (a).
Calculations are done with the crossover EoS.
}
\label{fig4}
\end{figure}
%
%
\begin{figure}[bht]
\includegraphics[width=7.cm]{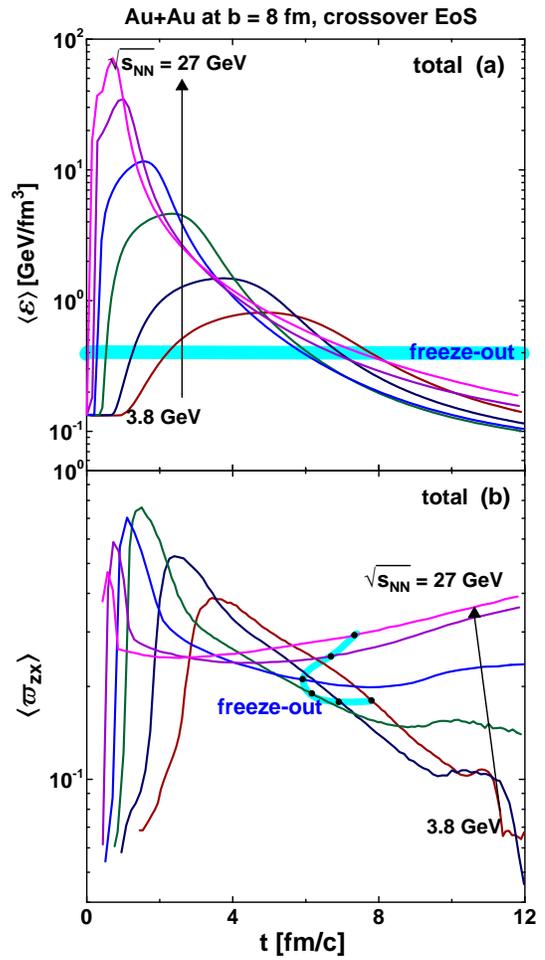}
 \caption{(Color online)
The same as in Fig \ref{fig4} but for averaging over the whole participant region  
at  collision energies $\sqrt{s_{NN}}=$ 3.8, 4.9, 7.7, 11.5, 19.6, 27 GeV. 
}
\label{fig5}
\end{figure}

Time evolution of the average energy density in the central slab and in 
the whole participant region is displayed in panels (a) of Figs. \ref{fig4} and \ref{fig5}, 
respectively. The bold cyan line indicates the freeze-out value $\varepsilon_{\rm frz}$. 
The simulations were performed with crossover EoS. 
We do not present results for the first-order-transition EoS because they are quite similar. 
As seen from Figs. \ref{fig4} and \ref{fig5}, the time span prior this global freeze-out is 
quite short. It should be compared to time of completion of the 
conventional local 3FD freeze-out at the same collision energies: 
8 fm/c for both total and central-slab freeze-out at 7.7 GeV,  
6 fm/c for central-slab freeze-out and 20 fm/c for total freeze-out at 39 GeV. 
This happens because the $\langle \varepsilon (t) \rangle$ value is calculated over 
all regions of the system, i.e. including those which would be already locally frozen out 
to the considered time instant.

The time evolution of the average proper-energy-weighted thermal $zx$ vorticity in 
the central slab and in 
the whole participant region is displayed in panels (b) of Figs. \ref{fig4} and \ref{fig5}, 
respectively. The bold cyan line indicates the global freeze-out which correspond to
the similar lines at value $\varepsilon_{\rm frz}$ in panels (a) of Figs. \ref{fig4} and \ref{fig5}. 

The central-slab thermal vorticity  rapidly decreases with time. 
At the early stages it practically coincides with the total one 
because this central region includes all the participant region. Later on   
the central vorticity becomes  
an order of magnitude and more lower than the total one. 
The central-slab vorticity at the freeze-out 
decreases with increasing collision energy because 
the vortical field is pushed out
to the fragmentation of regions \cite{Ivanov:2018eej,Jiang:2016woz}. 
The violation of this trend at energies 19.6 and 39 GeV is because of somewhat unstable numerics at
39-GeV energy.

At the same time, the average total thermal vorticity at the freeze-out generally rises 
with the collision energy 
as it can be expected from the 
corresponding increase of the total angular momentum accumulated in the participants, see Fig. \ref{fig1}. 
At the lowest considered energy of 3.8 GeV the central-slab and total values of the vorticity are 
very similar because the vorticity is more homogeneously distributed over the beam direction 
\cite{Kolomeitsev:2018svb,Ivanov:2017dff,Csernai:2014ywa} than at higher collision energies. 
The average total thermal vorticity as a function of time changes much slower as compared 
with the central one: at lower collision energies it moderately decreases while at higher energies  
even slightly rises with time.

\section{Polarization}
\label{polarization}


In terms of the mean spin vector (\ref{xp-pol}), 
the polarization vector of $S$-spin particle is defined as 
   \begin{eqnarray}
   \label{P_S}
  P^\mu_{S} = S^\mu / S.  
   \end{eqnarray}
In the experiment, the  polarization of the $\Lambda$ hyperon is measured in
its rest frame, therefore the $\Lambda$ polarization is 

   \begin{eqnarray}
   \label{P_L-rest}
  P^\mu_{\Lambda} = 2 S^{*\mu}_{\Lambda}  
   \end{eqnarray}
where $S^{*\mu}_{\Lambda}$ is mean spin vector of the $\Lambda$ hyperon in its rest frame. 
In the $\Lambda$ rest frame 
 the zeroth component  $S^{0}_{\Lambda}$
 identically vanishes and the spatial component becomes \cite{Kolomeitsev:2018svb}
   \begin{eqnarray}
\label{S-rest}
 {\bf S}^*_{\Lambda}(x,p)
 = {\bf S}_{\Lambda} - 
 \frac{{\bf p}_{\Lambda} \cdot {\bf S}_{\Lambda}}{E_{\Lambda}(E_{\Lambda}+m_{\Lambda})}
 {\bf p}_{\Lambda}. 
   \end{eqnarray}
Substitution of the  expression for ${\bf S}$ from Eq. (\ref{xp-pol}) and averaging this 
expression over the ${\bf p}_{\Lambda}$ direction (i.e. over ${\bf n}_p$) 
results in the following 
polarization in the direction orthogonal to the reaction plane ($xz$) \cite{Kolomeitsev:2018svb}
(see also \cite{Becattini:2013fla,Fang:2016vpj,Becattini:2016gvu}) 
   \begin{eqnarray}
   \label{P_Lambda}
\langle  P_{\Lambda}\rangle_{{\bf n}_p} =  
 \frac{1}{2m_{\Lambda}}
 \left(E_{\Lambda} - \frac{1}{3} \frac{{\bf p}_{\Lambda}^2}{E_{\Lambda}+m_{\Lambda}} \right)
 \varpi_{zx},  
   \end{eqnarray}
where $m_{\Lambda}$  is the $\Lambda$ mass, 
$E_{\Lambda}$ and ${\bf p}_{\Lambda}$ are the energy and momentum of the emitted $\Lambda$ hyperon, respectively. 
Here we put $(1-n_\Lambda) \simeq 1$ because the $\Lambda$ production takes place only 
in high-temperature regions, where Boltzmann statistics dominates.

Particles are produced across entire
freeze-out hypersurface. Therefore to calculate the global
polarization vector, the above
expression should be integrated over the freeze-out hypersurface $\Sigma$
and particle momenta 
   \begin{eqnarray}
\label{polint}
 \langle  P_{\Lambda}\rangle
 = \frac{\int (d^3 p/p^0) \int_\Sigma d \Sigma_\lambda p^\lambda
n_{\Lambda}  P_{\Lambda}}
 {\int (d^3 p/p^0) \int_\Sigma d\Sigma_\lambda p^\lambda \, n_{\Lambda}}. 
   \end{eqnarray}
Because of the isochronous freeze-out $(d^3 p/p^0) d\Sigma_\lambda p^\lambda =d^3 p \;d^3 x $.

\begin{figure}[bht]
\includegraphics[width=8.0cm]{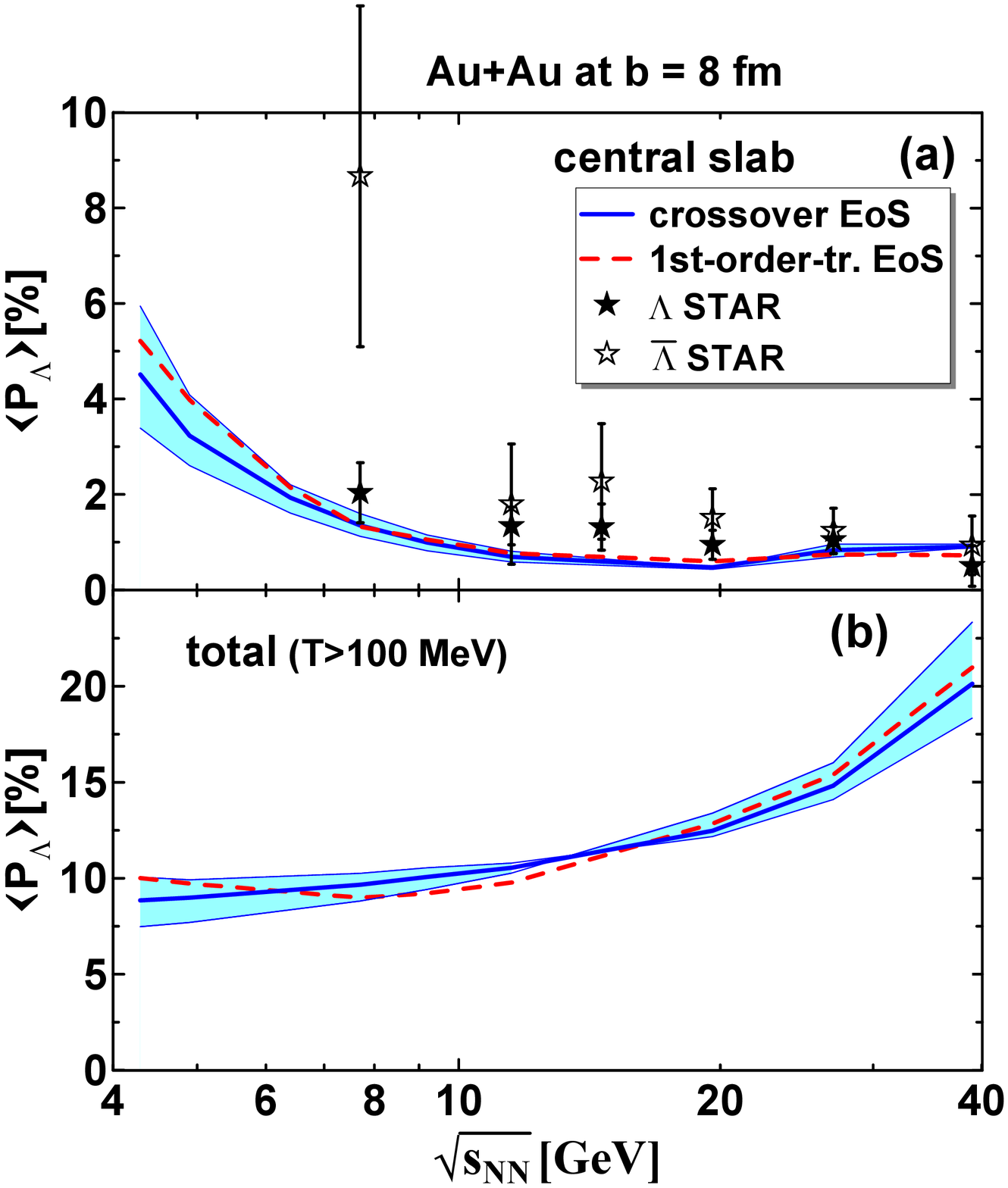}
 \caption{(Color online)
 Global (a), i.e. in the central-slab region, and total (b), i.e. averaged over the whole participant region,
polarization of $\Lambda$ hyperons in Au+Au collisions at $b=$ 8 fm as a function of collision energy $\sqrt{s_{NN}}$.  
The blue bands indicate polarization uncertainty due to a change the freeze-out 
criterion from  $\varepsilon_{\rm frz}$ = 0.3 to 0.5 GeV/fm$^3$ for the crossover EoS.  
STAR data on global $\Lambda$ and also $\bar{\Lambda}$ polarization in the midrapidity 
region (pseudorapidity cut $|\eta|<$ 1) \cite{STAR:2017ckg} are also displayed. 
}
\label{fig6}
\end{figure}

We apply further approximations after which  the present evaluation of the global
polarization becomes more an estimation rather than a calculation. 
We associate the global midrapidity polarization with the polarization of $\Lambda$ hyperons 
emitted from the above discussed central slab. 
We decouple averaging of $\varpi_{zx}$ and the term in parentheses in Eq. (\ref{P_Lambda}). 
Here we neglecte the longitudinal motion of the $\Lambda$ hyperon  
at the freeze-out stage in the central slab
and therefore approximate the average $\Lambda$ energy by the mean midrapidity transverse mass: 
$\langle E_{\Lambda}\rangle = \langle m_T^\Lambda \rangle_{\scr{midrap.}}$,
which was calculated earlier in Ref. \cite{Ivanov:2013yla}. 
Applying all the above approximations, we arrive at the  
 estimate of the global 
midrapidity $\Lambda$-polarization in the direction orthogonal to the reaction plane ($xz$) 
   \begin{eqnarray}
   \label{mid.r.-P_Lambda}
  \langle P_{\Lambda} \rangle_{\scr{midrap.}} \simeq 
 \frac{\langle \varpi_{zx} \rangle_{\scr{cent. slab}} }{2}
 \left(1 +  \frac{2}{3} 
 \frac{\langle m_T^\Lambda \rangle_{\scr{midrap.}} - m_\Lambda}{m_{\Lambda}} \right) 
\cr
   \end{eqnarray}
Results of this estimate are presented in panel (a) of Fig. \ref{fig6}. 
The corresponding 3FD simulations of Au+Au collisions were performed at fixed
impact parameters $b=$ 8 fm. This value of $b$ was chosen in order to roughly 
comply with the centrality selection 20-50\% in the STAR experiment \cite{STAR:2017ckg}. 
The correspondence between experimental centrality and the mean impact parameter
was taken from Glauber simulations of Ref. \cite{Abelev:2008ab}.

%
As seen from Fig. \ref{fig6}, such a rough estimate of the global midrapidity polarization
quite satisfactorily reproduces the experimental data, especially the collision-energy 
dependence of the polarization. This energy dependence is related to the decrease 
of the thermal vorticity in the central region (see Fig. \ref{fig4}) with the collision 
energy rise. The latter is a consequence of pushing out the vorticity field into  
the fragmentation region, which was discussed in Ref. \cite{Ivanov:2018eej} in detail. 
This effect of pushing out was found already in Ref. \cite{Jiang:2016woz}. 
Difference between results of the first-order-phase-transition and 
crossover EoS's is negligible. Apparently this is related to the fact that these two 
EoS's equally well reproduce the bulk of the available experimental data in this energy
range. The performed estimate predicts that the global midrapidity polarization 
further increases at NICA/FAIR energies, reaching values of 5\% at $\sqrt{s_{NN}}=$ 3.8 GeV. 
This prediction approximately agrees with that made in Ref. \cite{Baznat:2017jfj} 
based on the axial vortical effect \cite{Rogachevsky:2010ys,Gao:2012ix,Sorin:2016smp}.

The global midrapidity polarization of $\bar{\Lambda}$ hyperons differs only with  
replacement of $\langle m_T^{\Lambda} \rangle_{\scr{midrap.}}$ by 
$\langle m_T^{\bar{\Lambda}} \rangle_{\scr{midrap.}}$  
in Eq. (\ref{mid.r.-P_Lambda}) from that for $\Lambda$'s
and quantitatively does not exceed 5\% of that for $\Lambda$ hyperons. Therefore, 
we do not display it in Fig. \ref{fig6}. 

As mentioned above, the freeze-out applied in this calculation differes from that 
used in previous studies of the bulk and flow observables. We 
studied sensitivity of the 
polarization to a change of the freeze-out criterion that indirectly simulates 
the effect of different freeze-out procedures. The results of the change of the freeze-out
energy density  from  $\varepsilon_{\rm frz}$ = 0.3 to 0.5 GeV/fm$^3$ for 
the calculations with the crossover EoS are presented in Fig. \ref{fig6}.  
The lower $\varepsilon_{\rm frz}$ corresponds to the lower border of the displayed band. 
As seen from Fig. \ref{fig6}, the resulting variation of the  central-slab 
polarization gradually changes from 30\% at the energy of 4.3 GeV to 5\% at
39 GeV. Results for the 1st-order-transition EoS are similar. 

To further estimate uncertainties of the present estimation we performed calculations 
in a considerably smaller central box: $|x| < $ 2 fm, $|y| <$ 2 fm
and $|z| <$ 2 fm$/\gamma_{cm}$, i.e. with the box used in Ref. \cite{Arsene:2006vf}
to estimate densities achieved in the center of colliding nuclei. 
The difference from results in the used central slab in 
the above case depends on the collision energy but generally does not exceed 20\%. 
Another source of uncertainty is feed-down contribution due to 
decays of higher-mass hyperons,  
which are not included in the present
estimate. According to Refs. \cite{Karpenko:2016jyx,Li:2017slc,Becattini:2016gvu}, 
including $\Lambda$'s
from resonance decays reduces the $\Lambda$ polarization by 15\% to 20\%.
Though, the resonance decays increase the $\Lambda$ polarization by approximately 20\%
according to Ref. \cite{Kolomeitsev:2018svb}.

In the case of total $\Lambda$ polarization the integration in Eq. (\ref{P_Lambda}) 
runs over the whole participant 
range confined by the condition $T>T_0$ with $T_0=$ 100 MeV. In such averaging the above applied 
 decoupling of averaging of $\varpi_{zx}$ and the term in parentheses in Eq. (\ref{P_Lambda}) 
is even less justified than in the central slab. Therefore, we do 
even more rough estimate of the
mean total polarization of emitted $\Lambda$ hyperons 
%
   \begin{eqnarray}
   \label{mean-P_Lambda}
  \langle P_{\Lambda} \rangle_{\rm total} \approx  
 \frac{\langle \varpi_{zx} \rangle_{T>T_0}}{2}
     \end{eqnarray}
by neglecting the term in parentheses in Eq. (\ref{P_Lambda}). Note that this term 
is a correction, though not a negligible one. Sometimes it results in 
30\% correction for the central-slab 
polarization. Thus, this is another point to above-discussed list of uncertainties.

Results of this estimate of the total $\Lambda$ polarization
are presented in panel (b) of Fig. \ref{fig6}. 
The total $\Lambda$ polarization increases with collision energy rise. 
This is in contrast to the energy dependence of the midrapidity polarization. 
This increase is quite moderate as compared with the rapid rise of the 
angular momentum accumulated in the participant region, see Fig. \ref{fig1}. 
%

A peculiar feature is seen in Fig. \ref{fig6}(b). The lower and upper borders of the band,  
corresponding to lower and higher freeze-out
energy densities,  $\varepsilon_{\rm frz}$ = 0.3 and 0.5 GeV/fm$^3$, respectively, 
at low collision energies $\sqrt{s_{NN}}\leq$ 11.5 GeV, 
cross and then change their places at high collision energies. 
Thus, the total polarization rises with decrease of the freeze-out energy density at high 
collision energies. This can be expected from the evolution of the thermal vorticity 
displayed in Fig. \ref{fig5}. 
This observation indirectly indicates that 
the $\Lambda$ polarization in the fragmentation regions 
reaches high values at high collision energies.
Indeed, the fragmentation regions become dominant 
at later time instants because of their 
longer evolution (as compared to the central region) due to relativistic time dilation
caused by their high-speed motion with respect to the
central region.
Therefore, at late freeze-out, i.e. at lower $\varepsilon_{\rm frz}$, 
we see a larger relative contribution from the fragmentation regions 
in the total polarization than that at the early freeze-out. 
The increase of the total polarization with simultaneous decrease of the 
midrapidity one additionally confirms the conjecture on high values of 
the $\Lambda$ polarization reached in the fragmentation regions 
at high collision energies. 

In view of high degree of the polarization and therefore large values of $\varpi_{zx}$ 
the expansion of the exponential function in terms of $\varpi$ is definitely inapplicable
[see Eqs. (34) and (35) in Ref. \cite{Becattini:2013fla}]. 
Let us remind, that this expansion was used in deriving formula for the polarization in  
\cite{Becattini:2013fla}. This is another source of uncertainty of the present estimate of 
the total polarization at high energies.

At lower collision energies values of the total and midrapidity polarization are very close 
to each other, which reflects a more homogeneous distribution of the vortical field over 
the bulk of the produced matter. This spread into the bulk
is an effect of dissipation (or the shear viscosity in terms of the
conventional hydrodynamics). In the 3FD dynamics
it is a result of the 3FD dissipation which increases with collision-energy decrease 
\cite{Ivanov:2016vkw}.

\section{Summary}
\label{Summary}

We estimated the global polarization of $\Lambda$ 
and $\bar{\Lambda}$ hyperons in Au+Au collisions 
in the midrapidity region and the total polarization, 
i.e. averaged over all rapidities. This estimate was 
based on the thermodynamical approach 
\cite{Becattini:2013fla,Fang:2016vpj,Becattini:2016gvu}.    
The relevant vorticity was simulated within the 3FD model \cite{3FD}.  
Collision-energy scan in the energy range of FAIR, NICA and BES-RHIC was 
performed. The midrapidity results were compared with STAR data 
\cite{STAR:2017ckg}.

It is found that without any adjustment of the model parameters
the performed rough estimate of the global midrapidity polarization
quite satisfactorily reproduces the experimental STAR data on the $\Lambda$ polarization, 
especially the collision-energy 
dependence of the polarization. This energy dependence is a consequence of the decrease 
of the thermal vorticity in the central region with the collision 
energy rise, which in its turn results from pushing out the vorticity field into  
the fragmentation regions \cite{Ivanov:2018eej,Jiang:2016woz}. 
Difference between results of the first-order-phase-transition and 
crossover EoS's is negligible. Apparently this is related to the fact that these two 
EoS's equally well reproduce the bulk of the available experimental data in this energy
range. The performed estimate predicts that the global midrapidity polarization 
further increases at NICA/FAIR energies, reaching values of 5\% at $\sqrt{s_{NN}}=$ 3.8 GeV. 
This prediction approximately agrees with that made in Ref. \cite{Baznat:2017jfj} 
based on the axial vortical effect \cite{Rogachevsky:2010ys,Gao:2012ix,Sorin:2016smp}.

The global midrapidity polarizations of $\bar{\Lambda}$'s and $\Lambda$'s practically 
do not differ from each other within the present estimate. This is also true for all other 
hydrodynamic \cite{Karpenko:2016jyx,Xie:2017upb} and 
kinetic \cite{Li:2017slc,Sun:2017xhx,Kolomeitsev:2018svb,Wei:2018zfb,Shi:2017wpk}
calculations based on the thermodynamical approach. It is not quite clear whether 
this contradicts to the STAR data at energy of 7.7 GeV because of large error bars 
of the measured $\bar{\Lambda}$ polarization. However, there are approaches which 
naturally explain this difference. One of them is that directly based on the axial vortical effect 
\cite{Rogachevsky:2010ys,Gao:2012ix,Sorin:2016smp}. Application of this approach 
 within the quark-gluon string transport model \cite{Baznat:2017jfj} well reproduces 
both the $\bar{\Lambda}$ and $\Lambda$ polarizations and spliting between them. Another recently 
suggested approach \cite{Csernai:2018yok} based on a Walecka-like model can also 
explain the difference in the $\bar{\Lambda}$-$\Lambda$ polarizations. However, 
ability of this Walecka-like approach to describe absolute values of these 
polarizations still remains to be seen.

According to our estimate, 
the total $\Lambda$ polarization increases with collision energy rise,  
which is in contrast to the energy dependence of the midrapidity polarization. 
This increase is quite moderate compared to the rapid rise of the 
angular momentum accumulated in the participant region. 
The increase of the total polarization with simultaneous decrease of the 
midrapidity one suggests that  at high collision energies  
the fragmentation-region polarization reaches high values.


\begin{acknowledgments} 
Fruitful discussions with E. E. Kolomeitsev are gratefully acknowledged.
This work was carried out using computing resources of the federal collective usage center ``Complex for simulation and data processing for mega-science facilities'' at NRC "Kurchatov Institute", http://ckp.nrcki.ru/.
Y.B.I. was supported by the Russian Science
Foundation, Grant No. 17-12-01427, and the Russian Foundation for
Basic Research, Grants No. 18-02-40084 and No. 18-02-40085. 
A.A.S. was partially supported by  the Ministry of Education and Science of the Russian Federation within  
the Academic Excellence Project of 
the NRNU MEPhI under contract 
No. 02.A03.21.0005. 
\end{acknowledgments}


\begin{thebibliography}{999}
%
\bibitem{Liang:2004ph} 
  Z.~T.~Liang and X.~N.~Wang,
  Phys.\ Rev.\ Lett.\  {\bf 94}, 102301 (2005)
  Erratum: [Phys.\ Rev.\ Lett.\  {\bf 96}, 039901 (2006)]
  [nucl-th/0410079].
%
\bibitem{Betz:2007kg} 
  B.~Betz, M.~Gyulassy and G.~Torrieri,
  Phys.\ Rev.\ C {\bf 76}, 044901 (2007)
  [arXiv:0708.0035 [nucl-th]].
%
\bibitem{Gao:2007bc} 
  J.~H.~Gao, S.~W.~Chen, W.~t.~Deng, Z.~T.~Liang, Q.~Wang and X.~N.~Wang,
  Phys.\ Rev.\ C {\bf 77}, 044902 (2008)
  [arXiv:0710.2943 [nucl-th]].
%
\bibitem{STAR:2017ckg} 
  L.~Adamczyk {\it et al.} [STAR Collaboration],
  Nature {\bf 548}, 62 (2017)
  [arXiv:1701.06657 [nucl-ex]].
%
%
%
\bibitem{Karpenko:2016jyx} 
  I.~Karpenko and F.~Becattini,
  Eur.\ Phys.\ J.\ C {\bf 77}, no. 4, 213 (2017)
  [arXiv:1610.04717 [nucl-th]].
%
\bibitem{Xie:2017upb} 
  Y.~Xie, D.~Wang and L.~P.~Csernai,
  Phys.\ Rev.\ C {\bf 95}, no. 3, 031901 (2017)
  [arXiv:1703.03770 [nucl-th]].
%
\bibitem{Li:2017slc} 
  H.~Li, L.~G.~Pang, Q.~Wang and X.~L.~Xia,
  Phys.\ Rev.\ C {\bf 96}, no. 5, 054908 (2017)
  [arXiv:1704.01507 [nucl-th]].
%
\bibitem{Sun:2017xhx} 
  Y.~Sun and C.~M.~Ko,
  Phys.\ Rev.\ C {\bf 96}, no. 2, 024906 (2017)
  [arXiv:1706.09467 [nucl-th]].
%
\bibitem{Kolomeitsev:2018svb} 
  E.~E.~Kolomeitsev, V.~D.~Toneev and V.~Voronyuk,
  Phys.\ Rev.\ C {\bf 97}, no. 6, 064902 (2018)
  [arXiv:1801.07610 [nucl-th]].
%
\bibitem{Wei:2018zfb} 
  D.~X.~Wei, W.~T.~Deng and X.~G.~Huang,
  Phys.\ Rev.\ C {\bf 99}, no. 1, 014905 (2019)
  [arXiv:1810.00151 [nucl-th]].
%
\bibitem{Shi:2017wpk} 
  S.~Shi, K.~Li and J.~Liao,
  Phys.\ Lett.\ B {\bf 788}, 409 (2019)
  [arXiv:1712.00878 [nucl-th]].
%
\bibitem{Becattini:2013fla} 
  F.~Becattini, V.~Chandra, L.~Del Zanna and E.~Grossi,
  Annals Phys.\  {\bf 338}, 32 (2013)
  [arXiv:1303.3431 [nucl-th]].
%
\bibitem{Fang:2016vpj} 
  R.~h.~Fang, L.~g.~Pang, Q.~Wang and X.~n.~Wang,
  Phys.\ Rev.\ C {\bf 94}, no. 2, 024904 (2016)
  [arXiv:1604.04036 [nucl-th]].
%
\bibitem{Becattini:2016gvu} 
  F.~Becattini, I.~Karpenko, M.~Lisa, I.~Upsal and S.~Voloshin,
  Phys.\ Rev.\ C {\bf 95}, no. 5, 054902 (2017)
  [arXiv:1610.02506 [nucl-th]].
%
%
\bibitem{Rogachevsky:2010ys} 
  O.~Rogachevsky, A.~Sorin and O.~Teryaev,
  Phys.\ Rev.\ C {\bf 82}, 054910 (2010)
  [arXiv:1006.1331 [hep-ph]].
%
\bibitem{Gao:2012ix} 
  J.~H.~Gao, Z.~T.~Liang, S.~Pu, Q.~Wang and X.~N.~Wang,
  Phys.\ Rev.\ Lett.\  {\bf 109}, 232301 (2012)
  [arXiv:1203.0725 [hep-ph]].
%
\bibitem{Sorin:2016smp} 
  A.~Sorin and O.~Teryaev,
  Phys.\ Rev.\ C {\bf 95}, no. 1, 011902 (2017)
  [arXiv:1606.08398 [nucl-th]].
%
\bibitem{Baznat:2017jfj} 
  M.~Baznat, K.~Gudima, A.~Sorin and O.~Teryaev,
  Phys.\ Rev.\ C {\bf 97}, no. 4, 041902 (2018)
  [arXiv:1701.00923 [nucl-th]]. 
%
%
\bibitem{3FD}
 Yu. B. Ivanov, V. N. Russkikh, and V.D. Toneev,
 Phys. Rev. C {\bf 73}, 044904 (2006) 
[nucl-th/0503088].
%
%
\bibitem{Friman:2011zz} 
  B.~Friman, C.~Hohne, J.~Knoll, S.~Leupold, J.~Randrup, R.~Rapp and P.~Senger,
  Lect.\ Notes Phys.\  {\bf 814}, pp.1 (2011).
%
\bibitem{Kekelidze:2017ghu} 
  V.~D.~Kekelidze, V.~A.~Matveev, I.~N.~Meshkov, A.~S.~Sorin and G.~V.~Trubnikov,
  Phys.\ Part.\ Nucl.\  {\bf 48}, no. 5, 727 (2017).
%
  V.~D.~Kekelidze, R.~Lednicky, V.~A.~Matveev, I.~N.~Meshkov, A.~S.~Sorin and G.~V.~Trubnikov,
  Eur.\ Phys.\ J.\ A {\bf 52}, no. 8, 211 (2016).
%
\bibitem{Ivanov:2013wha} 
  Yu.~B.~Ivanov,
  Phys. Rev. C {\bf 87}, 064904 (2013) [arXiv:1302.5766 [nucl-th]]. 
%
\bibitem{Ivanov:2012bh} 
  Y.~B.~Ivanov,
  Phys.\ Lett.\ B {\bf 721}, 123 (2013)
  [arXiv:1211.2579 [hep-ph]];
%
  Y.~B.~Ivanov and D.~Blaschke,
  Phys.\ Rev.\ C {\bf 92}, no. 2, 024916 (2015)
  [arXiv:1504.03992 [nucl-th]].
%
\bibitem{Ivanov:2013yqa} 
  Y.~B.~Ivanov,
  Phys.\ Rev.\ C {\bf 87}, no. 6, 064905 (2013)
  [arXiv:1304.1638 [nucl-th]].
%
\bibitem{Ivanov:2013yla} 
  Y.~B.~Ivanov,
  Phys.\ Rev.\ C {\bf 89}, no. 2, 024903 (2014)
  [arXiv:1311.0109 [nucl-th]].
%
\bibitem{Ivanov:2014zqa} 
  Y.~B.~Ivanov and A.~A.~Soldatov,
  Phys.\ Rev.\ C {\bf 91}, no. 2, 024914 (2015)
  [arXiv:1401.2265 [nucl-th]];
%
  Y.~B.~Ivanov,
  Phys.\ Lett.\ B {\bf 723}, 475 (2013)
  [arXiv:1304.2307 [nucl-th]].
%
\bibitem{Konchakovski:2014gda} 
  V.~P.~Konchakovski, W.~Cassing, Y.~B.~Ivanov and V.~D.~Toneev,
  Phys.\ Rev.\ C {\bf 90}, no. 1, 014903 (2014)
  [arXiv:1404.2765 [nucl-th]];
%
  Y.~B.~Ivanov and A.~A.~Soldatov,
  Phys.\ Rev.\ C {\bf 91}, no. 2, 024915 (2015)
  [arXiv:1412.1669 [nucl-th]];
%
  Eur.\ Phys.\ J.\ A {\bf 52}, no. 1, 10 (2016)
  [arXiv:1601.03902 [nucl-th]].
%
\bibitem{Ivanov:2018vpw} 
  Y.~B.~Ivanov and A.~A.~Soldatov,
  Phys.\ Rev.\ C {\bf 97}, no. 2, 024908 (2018)
  [arXiv:1801.01764 [nucl-th]].
%
%
\bibitem{Adamczyk:2017iwn} 
  L.~Adamczyk {\it et al.} [STAR Collaboration],
  Phys.\ Rev.\ C {\bf 96}, no. 4, 044904 (2017)
  [arXiv:1701.07065 [nucl-ex]].
%
%
\bibitem{gasEOS}
V. M. Galitsky and I. N. Mishustin, Sov. J. Nucl. Phys. {\bf 29}, 181 (1979).
%
%
\bibitem{Toneev06}
A. S. Khvorostukhin,  
V. V. Skokov, K. Redlich, and V. D. Toneev,
Eur. Phys. J. {\bf C48}, 531 (2006)  [nucl-th/0605069].
%
\bibitem{Sat90} L. M.~Satarov, Sov. J. Nucl. Phys. {\bf 52}, 264 (1990).
%
%
%
\bibitem{Russkikh:2006aa} 
  V.~N.~Russkikh and Yu.~B.~Ivanov,
  Phys.\ Rev.\ C {\bf 76}, 054907 (2007)  [nucl-th/0611094];
%
  Yu.~B.~Ivanov and V.~N.~Russkikh,
  Phys.\ Atom.\ Nucl.\  {\bf 72}, 1238 (2009)  [arXiv:0810.2262 [nucl-th]].
%
%
%
\bibitem{Ivanov:2017dff} 
  Y.~B.~Ivanov and A.~A.~Soldatov,
  Phys.\ Rev.\ C {\bf 95}, no. 5, 054915 (2017)
  [arXiv:1701.01319 [nucl-th]].
%
\bibitem{Ivanov:2017xee} 
  Y.~B.~Ivanov and A.~A.~Soldatov,
  Phys.\ Rev.\ C {\bf 97}, no. 2, 021901 (2018)
  [arXiv:1711.03069 [nucl-th]];
%
  Phys.\ Rev.\ C {\bf 98}, no. 1, 014906 (2018)
  [arXiv:1803.11474 [nucl-th]].
%
\bibitem{Ivanov:2018eej} 
  Y.~B.~Ivanov and A.~A.~Soldatov,
  Phys.\ Rev.\ C {\bf 97}, no. 4, 044915 (2018)
  [arXiv:1803.01525 [nucl-th]].
%
\bibitem{Jiang:2016woz} 
  Y.~Jiang, Z.~W.~Lin and J.~Liao,
  Phys.\ Rev.\ C {\bf 94}, no. 4, 044910 (2016)
  Erratum: [Phys.\ Rev.\ C {\bf 95}, no. 4, 049904 (2017)]
  [arXiv:1602.06580 [hep-ph]].
%
\bibitem{Csernai:2014ywa} 
  L.~P.~Csernai, D.~J.~Wang, M.~Bleicher and H.~Stocker,
  Phys.\ Rev.\ C {\bf 90}, no. 2, 021904 (2014).
%
\bibitem{Abelev:2008ab} 
  B.~I.~Abelev {\it et al.} [STAR Collaboration],
  Phys.\ Rev.\ C {\bf 79}, 034909 (2009)
  [arXiv:0808.2041 [nucl-ex]].
%
\bibitem{Arsene:2006vf} 
  I.~C.~Arsene {\it et al.},
  Phys.\ Rev.\ C {\bf 75}, 034902 (2007)
  [nucl-th/0609042].
%
\bibitem{Ivanov:2016vkw} 
  Y.~B.~Ivanov and A.~A.~Soldatov,
  Eur.\ Phys.\ J.\ A {\bf 52}, no. 5, 117 (2016)
  [arXiv:1604.03261 [nucl-th]]; 
  Eur.\ Phys.\ J.\ A {\bf 52}, no. 12, 367 (2016)
  [arXiv:1605.02476 [nucl-th]].
%
\bibitem{Csernai:2018yok} 
L.~P.~Csernai, J.~I.~Kapusta and T.~Welle,
  Phys.\ Rev.\ C {\bf 99}, no. 2, 021901 (2019)
  [arXiv:1807.11521 [nucl-th]].
%
\end{thebibliography}
\end{document}